\documentclass[10pt,conference]{IEEEtran}
\usepackage{latexsym,graphics,graphicx,amssymb,amsmath, amstext, amsfonts, amsthm, pifont, color, subfig}

\newcommand{\nn}{\nonumber}
\newcommand{\bc}{\begin{center}}
\newcommand{\ec}{\end{center}}
\newcommand{\bfl}{\begin{flushleft}}
\newcommand{\efl}{\end{flushleft}}

\newcommand{\beqa}{\begin{eqnarray}}
\newcommand{\eeqa}{\end{eqnarray}}
\newcommand{\beqan}{\begin{eqnarray*}}
\newcommand{\eeqan}{\end{eqnarray*}}
\newcommand{\beq}{\begin{equation}}
\newcommand{\eeq}{\end{equation}}
\newcommand{\beit}{\begin{itemize}}
\newcommand{\eeit}{\end{itemize}}

\newcommand{\lbr}{\left \{ }
\newcommand{\rbr}{\right \} }
\newcommand{\Lbr}{\left [}
\newcommand{\Rbr}{\right ]}
\newcommand{\lp}{\left (}
\newcommand{\rp}{\right )}

\newtheorem{theorem}{Theorem}

\newcommand{\mc}{\mathcal}
\newcommand{\mb}{\mathbb}
\newcommand{\mbf}{\mathbf}

\makeatletter

\author{\authorblockN{Adnan Raja, Vinod M. Prabhakaran, and Pramod Viswanath}
\authorblockA{Coordinated Science Laboratory\\
University of Illinois, Urbana-Champaign, IL 61801\\
Email:\tt \{araja2, vinodmp, pramodv\}@uiuc.edu}}

\title{Reciprocity in Linear Deterministic Networks under Linear Coding}

\begin{document}
\maketitle

\begin{abstract}
The linear deterministic model has been used recently to get a first order understanding of many wireless communication network problems
\cite{ADT07}\cite{BPT07}\cite{BT08}\cite{MDFT08}. In many of these cases, it has been pointed out that the capacity regions of the network and its reciprocal (where the communication links are reversed and the roles of the sources and the destinations are swapped) are the same.
In this paper, we consider a linear deterministic communication network with multiple unicast information flows. For this model and under the restriction to the class of linear coding, we show that the rate regions for a network and its reciprocal are the same. This can be viewed as a generalization of the linear reversibility of wireline networks, already known in the network coding literature \cite{R08}.
\end{abstract}

\section{Introduction}\label{sec:intro}

We consider a network communication model $\mc{N}$ with multiple unicast flows.
The network is  represented by a directed graph
$\mc{G} =\lp \mc{V}, \mc{E} \rp$.  Here $\mc{V}$ is the set of vertices of the graphs representing the communication nodes
and $\mc{E}$ is the set of directed edges representing the communication links between the nodes.
Multiple unicast flows mean that there are $n$ independent messages {\em flowing} in the network.
Every message $i \in \lbr 1,\ldots,n \rbr$ is associated with a source-destination pair, $\mc{S}_{i}-\mc{D}_{i}$.
Note that two or more messages could have the same source node or the same destination node. A node could be a source for one message and a destination for the other.

In the standard wireless interaction model for the communication links \cite{TV05}, the transmitted signals $x_{i}[m]$ get attenuated by channel gains $h_{ij}$ to which independent Gaussian noise $z_{j}[m]$ is added to give the received signal $y_{j}[m]$. The corresponding complex baseband discrete-time model is:
\beq
y_j[m] = \sum_{i\in N_j} h_{ij} x_i[m] + z_j[m],
\label{eq:Gchannelmodel}
\eeq
where $N_{j}$ denotes the neighboring nodes and is the set of all nodes $i$, such that there is a directed edge from the node $i$ to the node $j$.
Here $y_{j}[m], x_{i}[m], h_{ij}$ and $z_{j}[m]$ are all complex numbers.

In \cite{ADT07} a simple {\em linear deterministic model} was introduced as a first step towards understanding the wireless interaction model.
The deterministic model simplifies the Gaussian wireless network model in \eqref{eq:Gchannelmodel} by eliminating noise and discretizing signal through a $p$-ary expansion.
Thus, at every time instant $m$, each node $i \in \mc{V}$ can transmit a signal which is a vector from a finite field $\mbf{x}_{i}[m] \in \mb{F}_{p}^{q}$. Here $\mb{F}_{p}$ is the finite field and $q$ is the length of the vector.
The channel gain is modeled by shift matrices $S_{ij}=S^{(q-g_{ij})}$, where
\beq
S = {\begin{bmatrix}
0 &0 &0 &\cdots &0 \\
1 &0 &0 &\cdots &0 \\
0 &1 &0 &\cdots &0 \\
\vdots &\ddots &\ddots &\ddots &\ddots \\
0 &\cdots &0 &1 &0
\end{bmatrix}}_{q\times q}
\eeq
and $g_{ij} \in \lbr 0,1,\cdots,q \rbr$ is a measure of the channel strength.
The shift matrices capture the attenuation effect of the signal caused by the channel gain by performing a shift on the $p$-ary representation of the input and ignoring the levels below the average noise level.
The superposition of signals is now modeled by finite field additions in $\mb{F}_{p}$. The received signal at any node can be written along the lines of \eqref{eq:Gchannelmodel} as,
\beq
\mbf{y}_{j}\Lbr m\Rbr = \sum_{i \in N_{j}} S_{ij} \mbf{x}_{i}\Lbr m\Rbr. \label{eq:SLDchannelmodel}
\eeq

This deterministic model was first proposed in the context of a compound point to point channel where it
was successfully used to construct codes that universally achieve the diversity-multiplexing tradeoff  
over any fading channel \cite{TV06} (see also Chapter~9 of \cite{TV05}). While this application handled
the broadcast nature of the wireless medium, it has been applied more recently to handle the interference
aspect of the wireless medium. Specifically: the 
 single source unicast and multicast problem \cite{ADT07-1}, the many-to-one and one-to-many interference channels \cite{BPT07}, and  a relay-interference network  \cite{MDFT08}. Formal connections between the deterministic model and  the corresponding Gaussian model have been explored in \cite{TV06} (see also Exercise 9.15 and 9.16 in \cite{TV05}) and in \cite{BT08}.

In this paper, we are interested in the relationship between a communication network and its reciprocal.
We define the reciprocal of a network as the network obtained by reversing the direction of the edges and swapping the roles of the source and destination nodes. To give two examples:
\beit
\item the many-to-one and the one-to-many interference channels in \cite{BPT07} are reciprocals of each other; and
\item the interference channel with transmitter cooperation and the interference channel with receiver cooperation
in \cite{PV08} are reciprocals of each other.
\eeit
In both these papers, a curious fact was pointed out: the capacity of the network and its reciprocal are the same.
Motivated by these facts, in this paper we investigate linear deterministic communication network with multiple unicast flows. Our main result is the following: {\em Under the class of linear coding schemes, the rates achievable for a given network and its reciprocal are the same.}

The rest of the paper is organized as follows. In section \ref{sec:LDN} the basic problem is setup and the main theorem on reciprocity is stated. In section \ref{sec:LLDN}, this theorem is proved for the special case of a layered network; we show separately how to convert any linear deterministic network into a layered one, this is handled in the appendix. Section \ref{sec:spcases} discusses the result for some special cases of interest.

\section{Main Results} \label{sec:main}

\subsection{Linear deterministic network} \label{sec:LDN}

For us, a linear deterministic communication network, denoted by $\mc{N}$, consists of
a) a directed graph $\mc{G}=(\mc{V},\mc{E})$, b) channel gain matrix $G_{ij}\in\mb{F}_{p}^{q\times q}$ associated with every edge $(i,j)\in \mc{E}$ and, c) $n$ unicast message flows with source-destination nodes denoted by $\mc{S}_{k}-\mc{D}_{k}$ and rate $R_{k}$, where $1\leq k\leq n$. 
The corresponding channel model is:
\beq
\mbf{y}_{j}\Lbr m\Rbr = \sum_{i \in N_{j}} G_{ij} \mbf{x}_{i}\Lbr m\Rbr. \label{eq:LDchannelmodel}
\eeq
If we restrict the channel matrices to shift matrices, we obtain the model described in \eqref{eq:SLDchannelmodel}.
We will call that the {\em shift linear deterministic network}.

We consider communication scheme over $T$ transmission times (symbols).
Every message $W_k$, for $1\leq k\leq n$, is a vector of independent symbols of length $w_kT$, i.e.~$W_k \in\mb{F}_{p}^{w_kT}$.
The message $W_k$ is available at the source node $\mc{S}_k$ and is demanded by the destination node $\mc{D}_k$.
The corresponding rate associated with the message $W_k$ is $R_k = w_k\log p$ bits. Thus the network is associated with a rate requirement $(R_1,\ldots,R_m)$.

At any time instant $m,\;0\leq m\leq T-1$, a node $j$ transmits a signal $x_j[m]$, which is a function, $f_j^{(m)}$, of the signals it has received so far $\lbr y_j[k]:\; 0\leq k\leq m-1\rbr$ and the source messages available at that node.
If $j$ is a destination node for the message with index $k$, it reconstructs an estimate $\hat{W}_k$, which is a function, $g_k$, of the received signals $\lbr y_j[k]:\; 0\leq k\leq T-1\rbr$.
A communication scheme is said to be  {\em linear coding}, if the functions $f_j^{(m)}$ and $g_k$ are linear.

We say that a {\em linear deterministic network} is {\em solvable} if there exists a coding scheme such that $\hat{W}_k=W_k$. Further if there exists a linear coding scheme, we say that the network is {\em linearly solvable}.
This follows the standard definitions in the (wireline) network coding literature \cite{DZF05}.

Given a network $\mc{N}$, its {\em reciprocal} is the network denoted by $\mc{N}^\prime$ and satisfying the following:
\beit
\item The nodes of $\mc{N}$ and $\mc{N}^\prime$ are the same.
\item The edges of $\mc{N}$ are $\mc{N}^\prime$ are the same, but their direction is reversed, i.e.,
$ (i,j) \in \mc{E} \Leftrightarrow (j,i) \in \mc{E}^{\prime}.$
\item The channel gain matrix associated with the same edges (but reversed) are transposes of each other, i.e.,
$G_{ji}^\prime = G_{ij}^T.$
\item The $n$ messages in $\mc{N}$ and $\mc{N}^\prime$ are the same but the role of the source and destination nodes
are interchanged.
\eeit
Note that for the shift deterministic network, the reciprocal is no longer a shift deterministic network, but rather a
{\em flipped} shift deterministic network, where the vectors are shifted upwards rather than downwards.
We discuss this  separately in section \ref{sec:spcases}.

We say that a network obeys {\em reciprocity} if the reciprocal of the network in solvable. Further, we say that a network obeys
{\em linear reciprocity} if the reciprocal of the network is linearly solvable. The main result of this paper is the following theorem.

\begin{theorem} \label{thm:1}
Any linear deterministic network which is linearly solvable obeys linear reciprocity.
\end{theorem}

Note that the above theorem is equivalent to the statement that {\em the linear deterministic network and its reciprocal have the same achievable rate region under the class of linear coding scheme}.
Our proof is broken up into two parts: we first prove the result for the special class of layered networks. Next
we show how, by unfolding over time, any linear deterministic network can be viewed as layered (without loss of any generality); this is done in an appendix.

\subsection{Layered linear deterministic network} \label{sec:LLDN}

\begin{figure}
\begin{center}
\scalebox{0.5}{\input{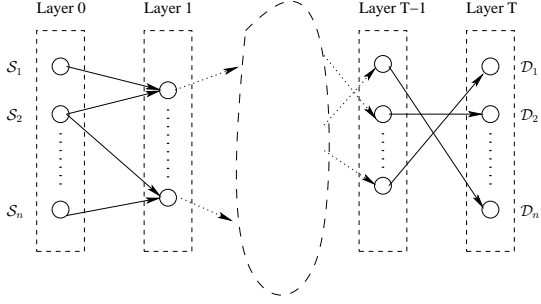}}
\end{center}
\caption{Layered network.}
\label{fig-layeredmodel}
\end{figure}

Consider a {\em $(T+1)$-layered} linear deterministic network illustrated in figure \ref{fig-layeredmodel}.
The layers are numbered $0$ to $T$, with the $0^{\mathrm{th}}$ layer comprising of source nodes and the $T^{\mathrm{th}}$ layer comprising of the destination nodes. The edges only connect nodes from one layer to the subsequent layer.
We consider a {\em layered} transmission scheme over such a network:
 each node only transmits once and it does so after it has received signals from the nodes in the previous layer.
Thus the source nodes transmit at time instant $0$. And the destination nodes eventually make a decision on the estimate of the message based on what they hear at time instant $T-1$.
The concepts of solvable, linearly solvable, reciprocity and linear reciprocity hold good for the layered network with the layered transmission scheme too.

In the appendix we show that any linear deterministic network described in section~\ref{sec:LDN} with a coding scheme over a block of time $T$
can be unfolded over time to create a layered network. Further, it is straightforward to see that the reciprocal of the original network corresponds to the reciprocal of the layered network. Thus it suffices to prove theorem \ref{thm:1}  for the layered network only. 

\begin{figure}
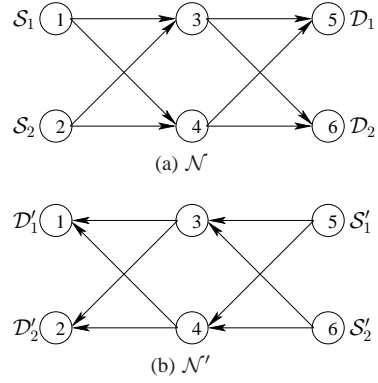

\begin{center}
\subfloat[$\mc{N}$]{\label{fig:Nmodel}\scalebox{0.6}{\input{figs/model4.pstex_t}}}\\
\subfloat[$\mc{N}^{\prime}$]{\label{fig:Npmodel}\scalebox{0.6}{\input{figs/model5.pstex_t}}}
\end{center}
\caption{An example of a Layered network with two-unicast (a);  and its reciprocal (b).}
\label{fig:egLmodel}
\end{figure}

\noindent{\em Proof of Theorem \ref{thm:1} (Layered networks):}
Consider a linear layered network $\mc{N}$, which is solvable by a linear coding scheme.
The linear coding scheme is specified by a set of linear matrices.
The coding matrices at the source nodes are denoted by $C_{k}\in \mb{F}_{p}^{q\times w_{k}T}$, for $k=1,\ldots,n$.
If $j$ is a source node for a subset of the messages, $\Omega_{j}\subseteq\lbr 1,\ldots, n\rbr$, then the signal transmitted by node $j$ is given by,
\beq x_{j} = \sum_{k\in\Omega_{j}} C_{k} W_{k}. \eeq
The decoding matrices are denoted by $D_{k}\in \mb{F}_{p}^{w_{k}T\times q}$, for $k=1,\ldots,n$.
If $j$ is a destination node for a subset of the messages, $\Omega_{j}\subseteq\lbr 1,\ldots,n\rbr$, then the node reconstructs the messages from the received signal $y_{j}$ using the decoding matrices,
\beq \hat{W}_{k} =  D_{k} y_{j},\quad\forall\;{k\in\Omega_{j}}. \eeq
Any intermediate relay node is associated with relay coding matrix $F_{j}$ such that,
\beq x_{j} = F_{j} y_{j}. \eeq
Note that since transmission happens at each node in the layered network only once, we can conveniently drop the time index.

The original messages at the source nodes, $\lbr {W}_{k} \rbr_{1}^{n}$,  and the reconstructed messages at the destination nodes, $\lbr \hat{W}_{k} \rbr_{1}^{n}$, can be related linearly as,
\beq
\hat{W}_{k} = \sum_{l} \Gamma_{lk} {W}_{l}.
\eeq
$\Gamma_{lk}$ is the transfer coefficient matrix between the source node $\mc{S}_{l}$ and the destination node $\mc{D}_{k}$.
It can be obtained by considering any path from the source node to the destination node, taking the product of the encoding and the channel matrices as you move along that path and then summing up this product for all such paths.
In the example of figure~\ref{fig:egLmodel}, the transfer coefficient matrices for the network $\mc{N}$ are
\begin{align}
\Gamma_{11} &= D_{1}G_{35}F_{3}G_{13}C_{1}+D_{1}G_{45}F_{4}G_{14}C_{1} \nn\\
\Gamma_{12} &= D_{2}G_{36}F_{3}G_{13}C_{1}+D_{2}G_{46}F_{4}G_{14}C_{1} \nn\\
\Gamma_{22} &= D_{2}G_{36}F_{3}G_{23}C_{2}+D_{2}G_{46}F_{4}G_{24}C_{2} \nn\\
\Gamma_{21} &= D_{1}G_{35}F_{3}G_{23}C_{2}+D_{1}G_{45}F_{4}G_{24}C_{2}. \nn
\end{align}
Since the network is solvable by this linear scheme, we must have $\hat{W}_{k}=W_{k}, \;\forall\;1\leq k\leq n$. Therefore, the transfer coefficient matrices $\Gamma_{lk}$ must satisfy the following condition:
\beq
\Gamma_{lk} = \delta_{lk}I. \label{eq:condLsolvability}
\eeq

Similarly, we can now consider the reciprocal network with the linear coding scheme parameterized by some coding matrices $C_{k}^{\prime}$, decoding matrices $D_{k}^{\prime}$ and relay coding matrices $F_{j}^{\prime}$. The transmitted messages $W_{k}$ and the reconstructed version $\hat{W}_{k}$ can again be linearly related by the matrices $\Gamma_{lk}^{\prime}$.
In the example of figure~\ref{fig:egLmodel}, the transfer coefficient matrices for the reciprocal network $\mc{N}^{\prime}$ are
\begin{align}
\Gamma_{11}^{\prime} &= D_{1}^{\prime}G_{13}^{T}F_{3}^{\prime}G_{35}^{T}C_{1}^{\prime}+D_{1}^{\prime}G_{14}^{T}F_{4}^{\prime}G_{45}^{T}C_{1}^{\prime} \nn\\
\Gamma_{12}^{\prime} &= D_{2}^{\prime}G_{23}^{T}F_{3}^{\prime}G_{35}^{T}C_{1}^{\prime}+D_{2}^{\prime}G_{24}^{T}F_{4}^{\prime}G_{45}^{T}C_{1}^{\prime} \nn\\
\Gamma_{22}^{\prime} &= D_{2}^{\prime}G_{23}^{T}F_{3}^{\prime}G_{36}^{T}C_{2}^{\prime}+D_{2}^{\prime}G_{24}^{T}F_{4}^{\prime}G_{46}^{T}C_{2}^{\prime} \nn\\
\Gamma_{21}^{\prime} &= D_{1}^{\prime}G_{13}^{T}F_{3}^{\prime}G_{36}^{T}C_{2}^{\prime}+D_{1}^{\prime}G_{14}^{T}F_{4}^{\prime}G_{46}^{T}C_{2}^{\prime}. \nn
\end{align}
Note that we have used the fact that $G_{ji}^{\prime}=G_{ij}^{T}$.
If we let,
\begin{align*}
C_{k}^{\prime}&=D_{k}^{T},\quad\forall\;1\leq k \leq n,\\
D_{k}^{\prime}&=C_{k}^{T},\quad\forall\;1\leq k \leq n,\\
\textrm{and}\quad F_{j}^{\prime}&=F_{j}^{T},
\end{align*}
then we can see that $\Gamma_{lk}^{\prime} = \Gamma_{kl}^{T}$.
Finally, from \eqref{eq:condLsolvability}, it follows that
\beq
\Gamma_{lk}^{\prime} = \delta_{lk}I.
\eeq
Therefore, the reciprocal network $\mc{N}^{\prime}$ is linearly solvable and hence the network $\mc{N}$ obeys linear reciprocity.
\hfill$\Box$

\subsection{Some special cases} \label{sec:spcases}

As mentioned earlier the general linear deterministic model of section~\ref{sec:LDN} captures two important special cases:
the noiseless wireline networks and the shift linear deterministic networks, which models the wireless networks.

\noindent{\em Noiseless wireline networks:}
Noiseless wireline networks have been extensively studied in network coding literature \cite{KM03} \cite{LNC03}.
They are characterized by orthogonal communication links, so that they are free of both features, interference and broadcast, present in
wireless networks. Our model captures the wireline networks as a special case obtained by choosing the channel gain matrices $G_{ij}$
such that the incoming and outgoing links become orthogonal.
The linear reciprocity result of theorem \ref{thm:1}, specialized to the wireline networks,
is already known \cite{R08}. The term reversibility is used in network coding literature instead of reciprocity.

\noindent{\em Shift linear deterministic networks:}
In section \ref{sec:intro} we introduced the linear deterministic network with the channel gain matrices being shift matrices $S_{ij} = S^{(q-g_{ij})}$. We saw that the shift matrix shifts a vector of length $q$ downwards by $q-g_{ij}$ levels and hence $g_{ij}$ represents the strength of the channel.
 Basic electromagnetic principles suggest that physical media are reciprocal, i.e. the communication link (channel) behaves the same way in both the forward and the reverse direction.
Therefore in the reciprocal network, we should expect the channel gain matrix to be the same.
However, in section \ref{sec:LDN} we defined the reciprocal of the network by taking the transpose of the channel gain matrices.
The transpose of the shift matrix $S_{ij}$ shifts the vector of length $q$ by $q-g_{ij}$ but instead of downwards this {\em shift is upwards}. An important observation here, is the following: the transpose of the shift matrix can be interpreted as a {\em flipping} of all the signal vectors.

More formally: consider the physical reciprocal network with the same channel gain matrix $S_{ij}$ on each edge as the original network. Given any coding scheme for this reciprocal network, we modify it as follows. Every node {\em flips} its vector before transmitting and {\em flips} the vector it receives before coding. The {\em flipping} operation is denoted by left-multiplying the vector with the matrix $J$, where
\beq
J = {\begin{bmatrix}
0 &\cdots &0 &0 &1 \\
0 &\cdots &0 &1 &0 \\
0 &\cdots &1 &0 &0 \\
\vdots &\ddots &\ddots &\ddots &\ddots \\
1 &0 &0 &\cdots &0
\end{bmatrix}}_{q\times q}.
\eeq
These matrices can be then ``absorbed'' into the channel matrix $S_{ij}$ to give the effective channel matrix $JS_{ij}J$. We readily see that $JS_{ij}J$ is same as $S_{ij}^{T}$.
In the context of the reciprocal network,  the actually encoding matrices for the {\em physical} reciprocal network should be
$JC_{k}^{\prime}J, JD_{k}^{\prime}J$ and $JF_{j}J$.

\appendix

\begin{figure}[ht]
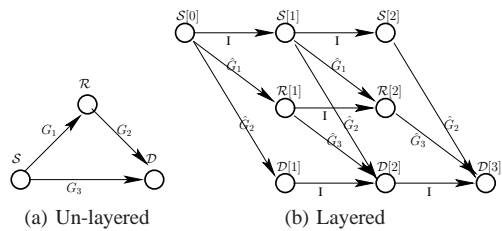

\begin{center}
\subfloat[Un-layered]{\label{fig:unlayered}\scalebox{0.35}{\input{figs/unlayered.pstex_t}}}
\subfloat[Layered]{\label{fig:layered}\scalebox{0.35}{\input{figs/layered.pstex_t}}}
\end{center}
\caption{Layering a network by unfolding over time.}
\label{fig:layering}
\end{figure}

\noindent {\em Layering Networks:}
We show that any linear deterministic network with a coding scheme (not necessarily linear) over $T$ time instants
can be unfolded over time to get an equivalent $T+1$ layered network with a layered coding scheme.
While the idea is similar to the one presented in \cite{ADT07}, our situation is  more general since we allow for coding schemes with arbitrary memory.
We unfold the network to $T+1$ stages such that the $m^{\mathrm{th}}$-stage is representing what happens in the network
during the  $m^{\mathrm{th}}$ time duration. Every node $\nu$ in the unlayered network appears at stage $1\leq m\leq T+1$ in the unfolded network as $\nu[m]$. If there is an edge connecting node $\nu_{i}$ to node $\nu_{j}$ with channel gain matrix $G_{ij} \in \mb{F}_{p}^{q\times q}$ in the unlayered network then there is an edge connecting the nodes $\nu_{i}[m]$ to node $\nu_{j}[m+1]$ in the layered network with channel gain matrix given by $\hat{G}_{ij}\in\mb{F}_{p}^{q(T+2)\times q(T+2)}$, where
\beq
\hat{G}_{ij} =
\begin{bmatrix}
0 & 0 & 0 \\
0 & 0 & 0 \\
G_{ij} & 0 & 0
\end{bmatrix}.
\eeq
Note that if $G_{ij}$ is a shift matrix then so is $\hat{G}_{ij}$.
Further, every node $\nu[m]$ is connected to its next instance $\nu[m+1]$ by a link with channel gain $I_{q(T+2)\times q(T+2)}$.
Figure \ref{fig:layering} illustrates a simple example of a network and the corresponding layered network.

We first show that any coding scheme for the unlayered network can be used to construct a layered coding scheme for the layered network.
If $x_{\nu_{j}}[m]\in\mb{F}_{p}^{q}$ is the vector transmitted by the node $\nu_{j}$ in the unlayered network, then the vector transmitted by the node $\nu_{j}[m]$ is $\hat{x}_{\nu_{j}[m]}\in\mb{F}_{p}^{q(T+2)}$ and is given by, 
\beq
\hat{x}_{\nu_{j}[m]} =
\begin{pmatrix}
x_{\nu_{j}}[m] \\
s_{\nu_{j}}[m-1] \\
0_{q\times 1}
\end{pmatrix},
\eeq
where $s_{\nu_{j}}[m-1]$ represents the state of the node $\nu_{j}$ at time instant $m-1$ in the unlayered network and is the stack of all received vectors at node $\nu_{j}$ until that time instant.
Note that the received vector at a node $\nu_{j}[m+1]$ in the layered network is 
\beq
\hat{y}_{\nu_{j}[m+1]} =
\begin{pmatrix}
x_{\nu_{j}}[m] \\
s_{\nu_{j}}[m-1] \\
y_{\nu_{j}}[m]
\end{pmatrix}.
\eeq
It is essentially all the information at node $\nu_{j}$ till time instant $m$. Thus any coding function for the node $\nu_{j}$ at time instant $m+1$ can be converted to a coding function for node $\nu_{j}[m+1]$.

Conversely, for any scheme on the layered network, the corresponding scheme on the unlayered network is given by, 
\beq x_{\nu_{j}}[m] = \lp \hat{x}_{\nu_{j}[m]}[1] \ldots \hat{x}_{\nu_{j}[m]}[q] \rp^{T}, \eeq
i.e.~$x_{\nu_{j}}[m]$ is the top $q$-part of the $\hat{x}_{\nu_{j}[m]}$.
It is easy to see that $x_{\nu_{j}}[m]$ can be written as a function of the previous received vectors and the source messages, if any, at that node.

\end{document}